\documentclass[aps,prc,preprintnumbers,twocolumn,nofootinbib]{revtex4-1}
\usepackage[dvips]{graphicx}
\usepackage{amsmath}
\usepackage{amssymb}
\usepackage{bm}
\newcommand{\bea}{\begin{eqnarray}}
\newcommand{\eea}{\end{eqnarray}}

\begin{document}
\preprint{arXiv: 1102.3752[nucl-th]}
\title{Effective range expansion in various scenarios of
EFT($\not\!\pi$)}
\author{Ji-Feng Yang$^{\dag,\ddagger}$}
\address{$^\dag$Department of Physics, East China Normal University,
Shanghai 200241, China\\
$^\ddagger$Kavli Institute for Theoretical Physics China, Chinese
Academy of Sciences, Beijing 100190, China}
\date{April 21, 2011}
\begin{abstract}
Using rigorous solutions, we compare the ERE parameters obtained in
three different scenarios of EFT($\not\!\!\pi$) in nonperturbative
regime. A scenario with unconventional power counting (like KSW) is
shown to be disfavored by the PSA data, while the one with elaborate
prescription of renormalization but keeping conventional power
counting intact seems more promising.
\end{abstract}
\maketitle
\section{Introduction}
Since Weinberg's seminal papers in the 1990s\cite{WeinEFT},
nuclear forces can now be pretty systematically understood and calculated
within the framework of effective field theories (EFT) basing on
chiral symmetry, for reviews on various achievements in this growing
area, we refer to Refs.\cite{Rev,oldrev}. In this course, there also
appear some intriguing issues. For example, according to the recent
review by Machleidt and Entem\cite{Mach}, the satisfactory
renormalization and power counting of the $NN$ sector within the
framework of EFT is still an open issue. There are roughly two main
types of choices towards this issue: One insists on the cut-off
independence with renormalization implemented through subtractions
according to new power counting schemes\cite{KSW,BKV,BBSvK,NTvK,
PVRA,EPVRA} that effectively invoke certain 'perturbative'
expansion around some leading nonperturbative components; The other
one stresses the nonperturbative tractability of the whole
approach\cite{EGM,EMach} with finite cutoff in the sense of
renormalization \'a la Lepage\cite{Lepage}. For approaches that do
not obviously fall in the above two choices, see
Refs.\cite{frederico,CJY,soto}. Recently, there appear new
evidences or arguments that seem to disfavor the first type choices
and the associated power counting schemes\cite{Epel-Gege,MLERA}.
Thus, this issue is closely tied with the rationality and
feasibility of various practical choices of the EFT power counting
of couplings and the prescription of renormalization in
nonperturbative regime. In this regard, it is much helpful to
explore this issue using closed-form $T$-matrices. As it is hard to
do so in presence of pion-exchanges, we turn to simplified situations
to gain useful hints.

In this report, we will explore the rationality of some typical
choices for power counting and renormalization prescriptions
through examining their predictions for the form factors in
effective range expansion (ERE) and confronting with the PSA
data\cite{PSA}, within the realm of EFT($\not\!\!\pi$)
whose renormalization has been clearly settled. This EFT is
employed due to its technical tractability and its practical
relevance to $NN$ scattering at lower momentum scales (say below
$70$MeV) and due to the fact that this EFT shares the same key
nonperturbative structures like power-like divergences and
truncation of potentials with the EFT containing pion-exchanges,
thus our studies here might be useful for the more general cases
with pion-exchanges. It will be shown that the choice with
unconventional power counting (like KSW) is strongly disfavored by
the PSA data. That means, in nonperturbative regime, modifying EFT
power counting would encounter more difficulties. Thus, such a
scheme is problematic in its rating of strengths of interactions
with or without pion exchanges, in line with other criticisms of
such schemes\cite{Epel-Gege,MLERA,Hansen-Cohen}. We feel such
discussions are worth doing as the EFT approaches are now widely
applied, see the recent application in systems like cold
atoms\cite{coldatom} where unnatural scales and nonperturbative
renormalization are also key issues.

This report is organized as follows: In Sec. II, the closed-form
solutions of $T$-matrices in EFT($\not\!\pi$) are presented,
where some issues related to renormalization are addressed and
clarified; Our main results and analysis of three typical
scenarios for EFT($\not\!\pi$) are presented in Sec. III. A
brief discussion and summary will be given in Sec. IV.
\section{Closed-form solutions and renormalization of EFT}
According to Weinberg, the potentials for $NN$ scattering (or
similar two-body non-relativistic scattering) are first
systematically constructed according to a reasonable set of power
counting rules for the EFT up to a given order $\Delta$ and then
resummed through Lippmann-Schwinger equations (LSE's) to obtain
the $T$-matrices\cite{WeinEFT}. Below pion threshold, the $NN$
potentials become contact ones (pionless EFT or EFT($\not\!\pi$)).
In an uncoupled channel with angular momentum $L$, we have:\bea
\label{factor}V_L(q,q^\prime)&&=\sum_{i,j=0}^{\Delta/2-L}C_{L;ij}
q^{2i+L}{q^\prime}^{2j+L}\nonumber\\&&=q^L{q^\prime}^LU^T(q^2)
C_{L}U({q^\prime}^2),\eea with $q,q^\prime$ being external momenta.
Introducing the column vector $U(x)\equiv(1,x,x^2,\cdots)$, the
potential is turned into a matrix sandwiched between vectors.
Note that: $C_{L;ij}=0,$ when $i+j>\Delta/2-L$ due to EFT
truncation. Here the couplings ${C}_{L;ij}$ are taken to be
energy-independent as the energy-dependence in potentials could
be removed using various methods\cite{E-dependence}. With such
contact potentials (EFT($\not\!\pi$)), one only needs to deal
with the following divergent integrals in Lippmann-Schwinger
equations:\bea{\mathcal{I}}_{ij}(E)\equiv\int\frac{d^3k}
{(2\pi)^3}\frac{k^{2(i+j)}}{E-k^2/M+i\epsilon},\ i,j\geq0.\eea
These integrals, which span a matrix ${\mathcal{I}}(E)\equiv
({\mathcal{I}}_{ij}(E))$, could be generally parametrized as
($p=\sqrt{ME}$):\bea\mathcal{I}_{ij}(E)\equiv\sum_{m=1}^{i+j}
J_{2m+1}p^{2(i+j-m)}-{\mathcal{I}}_0p^{2(i+j)},\eea with
${\mathcal{I}}_0\equiv J_0+i\frac{Mp}{4\pi}$ and the arbitrary
parameters [$J_0, J_{2m+1}, m>0$] parametrizing any sensible
prescription of regularization and/or renormalization.

Note that at any given order $\Delta$, the matrix ${\mathcal{I}}
(E)$ of finite rank characterizes the entire nonperturbative
structures of the divergences or ambiguities to be fixed, so only
finitely many divergences are involved, NOT infinitely many. This
feature should persists even in presence of pion exchanges.
With these preparations, we find that the on-shell $T$-matrix for
channel $L$ takes the following form at any given order of
truncation:\bea\label{Tinvuncouple}\frac{1}{T_L(p)}&=&\mathcal{I}
_0+\frac{{\mathcal{N}}_L}{{\mathcal{D}}_Lp^{2L}}.\eea The coupled
channels could be treated in the same fashion. The on-shell
$T$-matrices for the channels $^3\!S_1$-$^3\!D_1$ read:\bea
\label{TinvSD}\frac{1}{T_{ss}(p)}&=&\mathcal{I}_0+\frac{
{\mathcal{N}}_0+\mathcal{I}_0{\mathcal{N}}_1p^4}{{\mathcal{D}}_0
+\mathcal{I}_0{\mathcal{D}}_1p^4},\nonumber\\ \frac{1}{T_{dd}(p)}
&=&\mathcal{I}_0+\frac{{\mathcal{N}}_0+\mathcal{I}_0{\mathcal{D}}
_0}{[{\mathcal{N}}_1+\mathcal{I}_0{\mathcal{D}}_1]p^4},\eea In
whatever channels, the factors $[{\mathcal{N}}_{\cdots},
{\mathcal{D}}_{\cdots}]$ are real polynomials in terms of
$[C_{\cdots}]$, $[J_{2m+1}, m>0]$ and on-shell momentum $p$, the
concrete expressions will be given in a detailed
report\cite{detail}, where it will also be shown that the
relation ${\mathcal{D}}_{sd}^2+{\mathcal{D}}_1{\mathcal{N}}_0
={\mathcal{N}}_1{\mathcal{D}}_0$ in $^3\!S_1$-$^3\!D_1$. The $^1S_0$
case has been discussed in details in our previous work\cite{C71}.
Our analysis below will be basing on the solutions given in
Eqs.(\ref{Tinvuncouple},\ref{TinvSD}). To proceed, we define the
following notations:\bea\label{NDexp}\mathcal{N}_{\cdots}=\sum_{l=
0,1,\cdots}\mathcal{N}_{\cdots;l}p^{2l},\ \mathcal{D}_{\cdots}=
\sum_{l=0,1,\cdots}\mathcal{D}_{\cdots;l}p^{2l}.\eea We also note
that the factors $[{\mathcal{N}}_{\cdots},{\mathcal{D}}_{\cdots}]$
are independent of the parameter $\mathcal{I}_0$ or $J_0$, a fact
that is quite consequential\cite{C71,epl85,note}: Since the
functional dependence of the on-shell $T$'s upon $p$ is physical,
the nonperturbative compact form of $T$'s implies that $\mathcal{I}
_0$ is already physical or renormalization group invariant, no
longer a prescription-dependent parameter in contrast to the
perturbative regimes.

Obviously, the renormalization of the $T$-matrices in
Eqs.(\ref{Tinvuncouple},\ref{TinvSD}) could not be simply achieved
with conventional means without ruining their nonperturbative forms,
see Refs.\cite{C71,epl85,note} for a transparent demonstration. In
Ref.\cite{note} it is noted the intrinsic mismatch between the EFT
couplings and the nonperturbative divergences precludes the
conventional counter-terms from working: short of couplings (due to
truncation) for the 'unmatched' divergences. Consequently,
subtractions have to be done at loop level, with some residual
constants becoming independent parameters to be determined through
additional physical boundaries, while in perturbative contexts each
divergence could be absorbed into the EFT couplings and makes the
couplings 'run'.

Among the two main choices mentioned above, the second one could
be roughly seen as one instance of subtraction at loop
level\cite{Gege99}: The integral cutoff is kept finite as an
independent parameter in addition and determined by fitting to data
rather than absorbed into the couplings. Here, our general
parametrization of the subtractions instead of a single cutoff is
feasible as only finitely many divergences are involved due to
truncation. In our view, this is the true source of tractability of
EFT in nonperturbative regime.

The origin of additional parameters could be seen in this way.
Suppose we work with an underlying theory where no divergence shows
up. In the EFT limit or projection, some of the operators would
show up in the EFT lagrangian at a given order, while the high
energy 'details' become 'regulators'. Part of the 'details' could
be reabsorbed into the EFT couplings they 'match' with, there would
also be 'unmatched' part due to truncation, which have to be
fixed as independent parameters in EFT. The good news is that, only
finitely many of such parameters would be involved due to
truncation\cite{C71,epl85,note}.

In fact, the EFT upper scales are physical in the sense that they
correspond to the thresholds of EFT expansion, not the ordinary
regularization scales to be removed later. For example, in
EFT($\not\!\pi$), $\Lambda\sim m_{\pi}$; In presence of pion
exchanges, $\Lambda\sim m_{\rho}$.

Evidently, within EFT($\not\!\pi$), physics are encoded in the
functional dependence of on-shell $T$'s upon on-shell momentum, or
more specifically in the parameter $\mathcal{I}_0$ and the ratios
$[\mathcal{N}_{\cdots;i}/\mathcal{N}_{\cdots;0}$, $\mathcal{D}
_{\cdots;j}/\mathcal{N}_{\cdots;0}]$ (c.f., Eq.(\ref{NDexp})),
which are rational functions in terms of $[C_{\cdots}]$ and
$[J_{\cdots}]$. Thus, a power counting scheme must conspire with
appropriate prescriptions to yield desired behaviors in EFT
approach. This requires the two sets of parameters be treated on
the same footing, giving rise to the concept of scenario.
Different scenarios (modulo equivalence) would give rise to
different physics. Therefore, for one specific system, only one
scenario (and its equivalents) of EFT would be correct or
sensible choice.

In general, EFT parameters could be functions of the EFT expansion
parameter $\epsilon (\equiv\frac{Q}{\Lambda})$ with $Q$ being an
ordinary EFT scale and $\Lambda$ the upper scale. For later
convenience, we introduce the following dimensionless parameters:
$C_{ij}=\frac{4\pi}{M}\frac{\tilde{c}_{ij}(\epsilon)}{2^{i+j}
\Lambda^{2i+2j+1}}, J_{2k+1}=\frac{M}{4\pi}\mu^{2k+1}\tilde{j}
_{2k+1}(\epsilon),$ with ${\mu}\sim Q$ being an ordinary
renormalization scale. In fact, the functional forms of $[\tilde{c}
_{ij}(\epsilon)]$ and $[\tilde{j}_{2k+1}(\epsilon)]$ embody the
detailed patterns of fine-tuning in a scenario. For the purpose of
demonstration below, it suffices to define the fine-tuning in a
scenario in terms of $[\tilde{c}_{ij}(\epsilon)]$ only: \bea\frac
{\tilde{c}_{ij}(\epsilon)}{|\tilde{c}_{ij;0}(\epsilon)|}=\pm1+
\mathcal{O}(\epsilon^\sigma),\ \sigma\in[1,\kappa],\eea where
$\tilde{c}_{ij;0}(\epsilon)$ denotes the leading term in the
$\epsilon$-expansion and $\kappa$ denotes the lowest exponent of
the higher order contributions to the ERE parameters measured in
terms of $\epsilon$-expansion. This is the guiding principle for
determining the fine tuning exponent $\sigma$ in the following
discussions.
\section{Various scenarios of EFT($\not\!\pi$)}
Below, we examine three typical scenarios of EFT. At this stage,
we remind that our formulation is applicable to all
non-relativistic dynamics governed by short-distance interactions,
where the working energy is well below the threshold of the
quanta mediating the short-distance interactions so that contact
potentials could be effectively employed to describe the physical
processes.
\subsection{Natural and unnatural scenarios}
Let us defines three typical scenarios:\bea\label{SCA}\text{A:}
&&\ \tilde{c}_{ij}\sim{\mathcal{O}}(1);\ \tilde{j}_{2k+1}\sim
{\mathcal{O}}(1);\ J_0\sim\frac{M}{4\pi}Q;\\ \label{SCB}\text{B:}
&&\ \tilde{c}_{ij}\sim\frac{{\mathcal{O}}(1)}{\epsilon^{i+j+1}};
\ \tilde{j}_{2k+1}\sim{\mathcal{O}}(1);\ J_0\sim\frac{M}{4\pi}Q;\\
\label{SCC}\text{C:}&&\ \tilde{c}_{ij}\sim{\mathcal{O}}(1);\
\tilde{j}_{2k+1}\sim{\mathcal{O}}(1);\ J_0\sim\frac{M}{4\pi}
\Lambda.\eea

Obviously, scenario A is defined with conventional EFT power
counting for couplings and usual renormalization prescription, it
will lead to natural ERE parameters, hence a natural scenario.
Scenario B comprises of an unconventional power counting of
couplings and a usual prescription, the couplings are unnaturally
large. It will indeed lead to unnatural scattering. In scenario C,
conventional power counting of couplings is preserved, and the
renormalization prescription is as usual, but the scale of $J_0(=
\text{Re}(\mathcal{I}_0))$ is chosen to scale differently from the
preceding two scenarios. This is because as a physical (RG
invariant) parameter\cite{C71,epl85,note} (see the discussions
below Eq.(\ref{NDexp})), $J_0$ could take values as large as
$M\Lambda/(4\pi)$ since the upper scale $\Lambda$ itself is a
physical parameter of normal size, NOT a regularization cutoff as
mentioned above. Thus, this scenario is 'natural' in the sense
that all the scales involved are naturally sized. But it is indeed
able to lead to unnatural scattering upon fine-tuning, see below.

The exponent $\kappa$ could be read off from the concrete
expressions of $1/T$. For example, in $^1S_0$, we have (at
$\Delta=4$),\bea&&\quad\text{Re}\left(4\pi/(M\Lambda T_{0}(p))
\right)\sim\nonumber\\\text{A}:&&\ \epsilon+\frac{1+o(\epsilon^3)
+\frac{p^2}{\Lambda^2}\epsilon^3{\mathcal{O}}(1+o(\epsilon^3))+
\cdots}{\tilde{c}_0+o(\epsilon^5)+\frac{p^2}{\Lambda^2}
{\mathcal{O}}(1+o(\epsilon^3))+\cdots};\nonumber\\\text{B}:&&\
\epsilon+\frac{1+o(\epsilon)+\frac{p^2}{\Lambda^2}{\mathcal{O}}
(1+o(\epsilon))+\cdots}{\tilde{c}_0+o(\epsilon)+\frac{p^2}
{\Lambda^2}\epsilon^{-2}{\mathcal{O}}(1+o(\epsilon))+\cdots};
\nonumber\\ \text{C}:&&\ 1+\frac{1+o(\epsilon^3)+\frac{p^2}
{\Lambda^2}\epsilon^3{\mathcal{O}}(1+o(\epsilon^3))+\cdots}
{\tilde{c}_0+o(\epsilon^5)+\frac{p^2}{\Lambda^2}{\mathcal{O}}
(1+o(\epsilon^3))+\cdots}.\nonumber\\ &&\kappa_{\text{A}}=3;\
\kappa_{\text{B}}=1;\ \kappa_{\text{C}}=3.\eea The related
details would be given in Ref.\cite{detail}.
\subsection{Primary and qualitative results}
Now, we perform the magnitude analysis of the ERE parameters in
$S$-waves in the three scenarios defined above using the
closed-form $T$-matrices obtained Sec. II with the contact
potentials truncated at order $\Delta=4$. To proceed,
the following primary fine-tunings of the lowest coupling
$\tilde{c}_{00}$ are considered:\bea\text{Tuning I:}&&\
\tilde{c}_{00}\sim-1+{\mathcal{O}}(\epsilon)\ (\text{scenario
A,C}),\nonumber\\&&\ \epsilon\tilde{c}_{00}\sim-1+{\mathcal{O}}
(\epsilon)\ (\text{scenario B});\nonumber\\ \text{Tuning II:}&&
\ \tilde{c}_{00}\sim-1+{\mathcal{O}}(\epsilon^2),\ (\text
{scenario A,C}).\nonumber\eea The results for $^3S_1$
are listed in Tables I with tuning I for scenario A, B and C.
In order to yield a scattering length of size $(\epsilon\Lambda
^{-1}\mathcal{O}(1+o(\epsilon))$ in scenario B, one should use
$\epsilon\tilde{c}_{00}\sim+1+\mathcal{O}(\epsilon)$ instead,
with the rest of ERE parameters being the same as in Table I.
Evidently, scenario A corresponds to systems with natural size
of ERE parameters while B and C to those with unnatural ones.
Moreover, scenario B seems to correspond to systems with more
unnatural scales than scenario C. The results of the $^1S_0$ case
are presented in Table II, where the tuning II is used in
scenario C to yield a much larger ($\epsilon^{-2}$) scattering
length in $^1S_0$, while tuning I is used in scenario A and B.
Due to $\kappa_{\text{B}}=1$, it does not makes sense to consider
tuning II in scenario B at all, and tuning II would lead
essentially the same scaling in scenario A as it is a natural
scenario.

Interestingly, the leading term for $r_e$ is very simple and
involve the coupling $\tilde{c}_{01}(=\tilde{c}_{10})$ only, so it
is listed out explicitly. Moreover, it is in fact of the same size
in the three scenarios as $\tilde{c}_{01}\sim\epsilon^{-2}$ in
scenario B. From $v_2$ on, the leading terms are found to involve
more and more couplings, leaving plenty room for further reduction
in size upon cancelation amongst the couplings, which will be
discussed in detail in the future\cite{detail}. Here, they are
simply assumed to be of order one to focus on our main concerns.
\begin{table}[h]\label{table1}\caption{Scaling of ERE parameters
in $^3S_1$}\begin{center}\begin{tabular}{|c|c|c|c|}
\hline\hline ERE & Scenario A & Scenario B & Scenario C\\
\hline $\Lambda\cdot a$& ${\mathcal{O}}(1+o(\epsilon))$ &
$\epsilon^{-2}\mathcal{O}(1+o(\epsilon))$&$\epsilon^{-1}
\mathcal{O}(1+o(\epsilon))$\\$\Lambda\cdot r_e$&$(2
\tilde{c}_{01}+o(\epsilon))$&$(2\epsilon^2\tilde{c}_{01}
+o(\epsilon))$&$(2\tilde{c}_{01}+o(\epsilon))$\\ $\Lambda^3
\cdot v_2$&${\mathcal{O}}(1+o(\epsilon))$&$\epsilon^{-1}
{\mathcal{O}}(1+o(\epsilon))$&${\mathcal{O}}(1+o(\epsilon))$\\
$\Lambda^5\cdot v_3$&${\mathcal{O}}(1+o(\epsilon))$&$\epsilon
^{-2}{\mathcal{O}}(1+o(\epsilon))$&${\mathcal{O}}(1
+o(\epsilon))$\\$\Lambda^7\cdot v_4$&${\mathcal{O}}(1
+o(\epsilon))$&$\epsilon^{-3}{\mathcal{O}}(1+o(\epsilon))$
&${\mathcal{O}}(1+o(\epsilon))$\\ \hline\hline\end{tabular}
\end{center}\end{table}
\begin{table}[h]\label{table1S0}
\caption{Scaling of ERE parameters in $^1S_0$}
\begin{center}\begin{tabular}{|c|c|c|c|}\hline\hline
ERE&Scenario A&Scenario B&Scenario C\\ \hline$\Lambda\cdot
a$&${\mathcal{O}}(1+o(\epsilon))$&$\epsilon^{-2}\mathcal{O}
(1+o(\epsilon))$&$\epsilon^{-2}\mathcal{O}(1+o(\epsilon^2))$
\\$\Lambda\cdot r_e$&$(2\tilde{c}_{01}+o(\epsilon^2))$
&$(2\epsilon^2\tilde{c}_{01}+o(\epsilon))$&$(2\tilde{c}_{01}
+o(\epsilon^2))$\\$\Lambda^{3}\cdot v_2$&${\mathcal{O}}(1
+o(\epsilon^2))$&$\epsilon^{-1}{\mathcal{O}}(1+o(\epsilon))$
&${\mathcal{O}}(1+o(\epsilon^2))$\\$\Lambda^{5}\cdot v_3$&
${\mathcal{O}}(1+o(\epsilon^2))$&$\epsilon^{-2}{\mathcal{O}}
(1+o(\epsilon))$&${\mathcal{O}}(1+o(\epsilon^2))$\\$\Lambda^{7}
\cdot v_4$ &${\mathcal{O}}(1+o(\epsilon^2))$&$\epsilon^{-3}
{\mathcal{O}}(1+o(\epsilon))$&${\mathcal{O}}(1+o(\epsilon^2))$
\\\hline\hline\end{tabular}\end{center}\end{table}

In order to see the rationality in choosing appropriate
prescriptions rather than modifying the canonical EFT power
counting, the empirical ERE parameters in $S$-waves are listed
and analyzed with respect to scaling using the PSA data that are
given in Table 1 of Ref.\cite{PSA}, and in Table 8 and Table 9
of Ref.\cite{epel747}. The results are given in Table III with
choice $\Lambda\approx m_{\pi^{\pm}}$ and $\epsilon\approx
\frac{1}{4}$. For example, the scattering lengths scale as
below: $\Lambda\cdot a_{^3S_1}\sim\epsilon^{-1},\ \Lambda
\cdot a_{^1S_0}\sim\epsilon^{-2}.$ From this table, one can see
that the PSA data lead to small $v_2$ in comparison to all the
three schemes above. In particular, the PSA data give an
extremely small $v_2$ in $^3S_1$ channel.
\begin{table}[h]\label{tablePSA}
\caption{PSA data and their scaling in $S$-waves}
\begin{center}\begin{tabular}{|c|c|c|c|c|}\hline\hline
ERE&$^3S_1$(data)&$^3S_1$(scaling)&$^1S_0$(data)&$^1S_0$
(scaling)\\ \hline $\Lambda\cdot a$&$(0.26)^{-1}$&$\epsilon^{-1}
{\mathcal{O}}(1)$&$-(0.06)^{-1}$&$\epsilon^{-2}{\mathcal{O}}(1)$\\
$\Lambda\cdot r_e$&$(0.81)^{-1}$&${\mathcal{O}}(1)$&$(0.53)^{-1}$
&$2{\mathcal{O}}(1)$\\$\Lambda^{3}\cdot v_2$&$(4.13)^{-3}$&$
\epsilon^3{\mathcal{O}}(1)$&$-(1.81)^{-3}$&$\epsilon^{\frac{5}{4}}
{\mathcal{O}}(1)$\\$\Lambda^{5}\cdot v_3$&$(1.53)^{-5}$&$\epsilon
^{\frac{3}{2}}{\mathcal{O}}(1)$&$(1.07)^{-5}$&${\mathcal{O}}(1)$\\
$\Lambda^{7}\cdot v_4$& $-(1.16)^{-7}$&$\epsilon^{\frac{3}{4}}
{\mathcal{O}}(1)$&$-(0.92)^{-7}$&${\mathcal{O}}(1)$\\ \hline\hline
\end{tabular}\end{center}\end{table}
\subsection{Scenarios and unnaturalness in ERE}
Let us elaborate on the scenarios B and C that lead to large
scattering lengths in $S$-waves. The magnitudes for $v_2, v_3$ and
$v_4$ obtained in scenario B seem to be quite large, contrary to
the PSA data that give much smaller numbers. In comparison, the
numbers given by scenario C are smaller and hence seem closer to
the PSA data. In $^1S_0$ channel, the agreement between scenario C
and PSA data are almost complete. Thus, the more complicated
scenario B seems to be disfavored in this regard.

If one requires that the PSA data be reproduced in the two
scenarios, then further fine-tunings are necessary. Comparing
Tables I and II with Table III, it is evident that there are huge
'gaps' in the size of ERE parameters between PSA data and scenario
B:\bea ^3S_1:&&\frac{v_{2;B}}{v_{2:P}}\sim\epsilon^{-4},\ \frac
{v_{3;B}}{v_{3;P}}\sim\epsilon^{-\frac{7}{2}},\ \frac{v_{4;B}}
{v_{4;P}}\sim\epsilon^{-\frac{15}{4}},\\^1S_0:&&\frac{v_{2;B}}
{v_{2:P}}\sim\epsilon^{-\frac{9}{4}},\ \frac{v_{3;B}}{v_{3;P}}
\sim\epsilon^{-2},\ \frac{v_{4;B}}{v_{4;P}}\sim\epsilon^{-3}.\eea
It seems extremely difficult to do the fine-tuning to remove these
'gaps' in scenario B. In contrast, the 'gaps' between PSA and
scenario C are much smaller in each ERE parameter:\bea\label{gaps}
&&^3S_1:\frac{v_{2;C}}{v_{2:P}}\sim\epsilon^{-3},\ \frac{v_{3;C}}
{v_{3:P}}\sim\epsilon^{-\frac{3}{2}},\ \frac{v_{4;C}}{v_{4:P}}\sim
\epsilon^{-\frac{3}{4}},\\&&^1S_0:\frac{v_{2;C}}{v_{2:P}}\sim
\epsilon^{-\frac{5}{4}},\ \frac{v_{3;C}}{v_{3:P}}\sim\epsilon^{0},
\ \frac{v_{4;C}}{v_{4:P}}\sim\epsilon^{0}.\eea Thus, scenario B is
also disfavored in the perspective of fine-tuning.

The problem with scenario B or unnatural couplings could also be
seen as follows: Suppose one insists on using the unconventional
couplings of scenario B to describe the $NN$ low-energy scattering
in terms of pionless EFT. Then to reproduce the scaling in scenario
C, it turns out that one has to choose the following scaling of the
renormalization parameters:\bea\frac{4\pi J_{2k+1}}{M\mu^{2k+1}}
\sim\epsilon^{k+1+a},\ a\geq0,\ \forall k>0.\eea which means that
$NN$ scattering with unnatural couplings would involve subtractions
at scales much smaller than normally expected, i.e., a very
'unnatural' prescription of renormalization. However, such
unconventional choice of renormalization could not be simply
excluded, though its rationales remain to be seen. Actually, in
the literature using modified power counting for couplings these
parameters were set to be even smaller: zero\cite{earlier}.

Here, we should remind again that our analysis are performed in
pionless situation. Since truncation is still necessary in cases
with pion exchanges, the basic 'characters' of the scenario issue
depicted here might remain, though it is more difficult to
work out closed-form solutions there. It would also be interesting
to see how the 'pictures' evolve after other contents of potential
are included\cite{soto-static}.

Although scenario C looks better than scenario B, it remains to
see how the further reduction of the 'gaps' in Eq.(\ref{gaps})
could happen. This issue will be addressed in the detailed
report\cite{detail}. In fact, closer studies may lead to more
constraints on the contact couplings in similar fashion using ERE
and/or other phenomenological data, the so-determined EFT
couplings may in turn provide useful targets for lattice studies
basing on more fundamental theories like QCD. Such kind of
analysis will also be given in our detailed report\cite{detail}.
\section{Discussions and summary}
Tables I, II and III are our main results of this report. As is
clear from the numbers listed above, the choice of employing
unconventional EFT power counting and conventional subtraction
for $NN$ scattering is disfavored by the PSA data. In other words,
modification of conventional power counting is disfavored in
comparison with choosing appropriate prescriptions in
nonperturbative regimes while keeping the conventional rating of
interactions intact. Of course, a conventional power counting of
couplings AND a conventional prescription could not be compatible
with unnatural scattering lengths. Thus the choice like scenario
C seems more promising, as our analysis done here is only a crude
one, there are still much rooms for further tuning to remove the
'gaps' as explained above. We will demonstrate instances of such
tuning in the detailed report\cite{detail}.

In the course of our presentation, it also occurs to us that EFT
truncation turns out to be an virtue rather than a burden in the
issue of renormalization of EFT in nonperturbative regime: It is
the truncation that keeps the number of nonperturbative
divergences involved, here the rank of the matrix ${\mathcal{I}}(E)$,
finite. Without truncation, the rank of ${\mathcal{I}}(E)$
would be infinite, an intractable situation that renders EFT
approach totally useless. This virtue might remain somehow in the
presence of pion exchanges and somewhat underlies the observation
that only finite nonperturbative subtractions are needed at a
given order\cite{NTvK,frederico,CJY}.
\section*{Acknowledgement}The author is grateful to Bira van
Kolck, Yu-Qi Chen, Yu Jia, Fan Wang, Antonio Pineda, Hong-Ying Jin,
Dao-Neng Gao and other participants at the KITPC program 'EFT's in
Particles and Nuclear Physics' (2009) for many helpful and
enlightening conversations over the EFT topics. Special thanks are
due to E. Epelbaum for his informative and helpful communications
over the issues discussed here. The author is also grateful to
the anonymous referees for their criticisms that improve the
presentation of our manuscript. The project is supported in part by
the Ministry of Education of China.

\end{document}